\documentclass[preprint]{aastex}
\usepackage{emulateapj5}
\usepackage{apjfonts}
\input psfig.tex

\def\aa{{A\&A}}

\def\aj{{AJ}}
\def\al{$\alpha$}

\def\apj{{ApJ}}
\def\apjs{{ApJS}}
\def\asec{$^{\prime\prime}$}

\def\deg{$^{\circ}$}

\def\etal{{et al. }}

\def\gax{{$\mathrel{\hbox{\rlap{\hbox{\lower4pt\hbox{$\sim$}}}\hbox{$>$}}}$}}

\def\hst{{\it HST}}

\def\lamb{$\lambda$}
\def\lax{{$\mathrel{\hbox{\rlap{\hbox{\lower4pt\hbox{$\sim$}}}\hbox{$<$}}}$}}

\def\nat{{Nature}}

\def\percm2{cm$^{-2}$}

\def\nii{[\ion{N}{2}]}

\received{}
\accepted{}
\slugcomment{To appear in {\it Publications of the Astronomical Society 
of the Pacific.}}
\shorttitle{BLACK HOLE MASS MEASUREMENTS USING HST}
\shortauthors{HO ET AL.}
\journalid{}{}
\articleid{}{}

\begin{document}

\title{An Efficient Strategy to Select Targets for Gas-Dynamical Measurements 
of Black Hole Masses Using the {\it Hubble Space Telescope}\footnotemark[1]}

\footnotetext[1]{Based on observations made with the {\it Hubble Space
Telescope}, which is operated by AURA, Inc., under NASA contract NAS5-26555.}

\author{
Luis C. Ho\altaffilmark{2}, 
Marc Sarzi\altaffilmark{3,4},
Hans-Walter Rix\altaffilmark{3},
Joseph C. Shields\altaffilmark{5},
Greg Rudnick\altaffilmark{6},
Alexei V. Filippenko\altaffilmark{7}, and
Aaron J. Barth\altaffilmark{8,9}
}

\altaffiltext{2}{The Observatories of the Carnegie Institution of Washington,
813 Santa Barbara St., Pasadena, CA 91101; lho@ociw.edu.}

\altaffiltext{3}{Max-Planck-Institut f\"{u}r Astronomie, K\"{o}nigstuhl 17,
Heidelberg, D-69117, Germany; rix@mpia-hd.mpg.de.}

\altaffiltext{4}{Dipartimento di Astronomia, Universit\`a di Padova,
  Vicolo dell'Osservatorio 5, I-35122 Padova, Italy; sarzi@pd.astro.it.}

\altaffiltext{5}{Physics \& Astronomy Department, Ohio University,
  Athens, OH 45701; shields@phy.ohiou.edu.}

\altaffiltext{6}{Steward Observatory, Univ. of Arizona, Tucson, 
AZ 85721; grudnick@as.arizona.edu.}

\altaffiltext{7}{Department of Astronomy, University of California, Berkeley,
CA 94720-3411; alex@astro.berkeley.edu.}

\altaffiltext{8}{Harvard-Smithsonian Center for Astrophysics, 60
  Garden Street, Cambridge, MA 02138; abarth@cfa.harvard.edu.}

\altaffiltext{9}{California Institute of Technology, Dept. of Astronomy, 
105-24, Pasadena, CA 91125.}

\begin{abstract}

Gas-dynamical studies using the {\it Hubble Space Telescope}\ are an integral 
component for future progress in the search for massive black holes in 
galactic nuclei.  Here we present an extensive set of gas rotation curves 
obtained with the Space Telescope Imaging Spectrograph for the central 
regions of 23 disk galaxies.  We find that the bulges of randomly selected, 
nearby spiral and S0 galaxies generally do not contain well-defined gaseous 
disks.  Only 15\%--20\% of disk galaxies have regular, symmetric velocity 
fields useful for dynamical analysis.  Through comparison of the kinematics 
with {\it Hubble Space Telescope}\ images of the nuclear regions, we show 
that the probability of success can be significantly boosted by preselecting 
objects whose central dust lanes follow a well-ordered, circularly symmetric 
pattern.  The dust morphology can be ascertained efficiently by visual 
inspection of unsharp-masked images.

\end{abstract}


\section{Introduction}

The installation of the Space Telescope Imaging Spectrograph (STIS; Leitherer 
et al. 2001) on the {\it Hubble Space Telescope (HST)}\ ushered in a new era 
in the search for massive black holes (BHs) in galactic nuclei.  By now, 
sufficiently large numbers of BH candidates are known that the emphasis has 
shifted from detections of individual objects to demographic studies using 
statistical samples (e.g., Richstone et al. 1998; Ho 1999; Gebhardt et al. 
2000b; Ferrarese \& Merritt 2000).  Future progress will focus on two fronts, 
first on sharpening the analysis techniques to improve the accuracy of the 
mass measurements, and second on substantially enlarging the existing samples 
to cover galaxies across the Hubble sequence.

BH searches with \hst\ use either stars or gas to trace the central potential.
These two approaches are complementary, each with its merits and limitations.
Modern, fully general, stellar-dynamical models are powerful because they give 
information not only on masses but also on the orbital structure of the galaxy 
(e.g., van~der~Marel et al. 1998; Cretton \& van~den~Bosch 1999; Gebhardt et 
al. 2000a).  One drawback is that both the data and the computational 
requirements are expensive. Another is that the modeling can be very 
challenging for complicated dynamical structures; the double nucleus of M31 
serves as a good illustration (e.g., Kormendy \& Bender 1999; Bacon et al. 
2001).  This technique also cannot easily cope with dusty galaxies, as a 
consequence of which most stellar-dynamical work with \hst\ has been biased 
toward early-type (mostly elliptical) galaxies.  

Gas kinematics in principle offer a more straightforward measure of the 
central mass, provided that the gas participates in Keplerian rotation in a 
disklike configuration.  The observations are less demanding because optical 
emission lines from ionized gas generally have much higher equivalent widths 
than stellar absorption lines.  Moreover, as demonstrated later in this paper, 
optical nebular emission invariably traces nuclear dust features.  This 
technique, therefore, can be applied to galaxies inaccessible to stellar 
observations.  Unlike stars, however, gas behaves as a collisional fluid and 
responds to nongravitational perturbations such as shocks, radiation pressure, 
or magnetic fields. Despite these potential complications, gas-dynamical 
models have been applied to a number of elliptical galaxies with 
circumnuclear gas disks (Harms et al. 1994; Ferrarese, Ford, \& Jaffe 1996; 
Macchetto et al. 1997; van~der~Marel \& van~den~Bosch 1998; Bower et al. 
1998; Ferrarese \& Ford 1999; Verdoes~Kleijn et al. 2000).
 
Our group has concentrated on using STIS to apply the gas-dynamical 
technique to the bulges of spiral and S0 galaxies (Shields et al. 2000; 
Barth et al. 2001; Sarzi et al. 2001, 2002).  Our work has revealed a 
subtlety that has not been widely appreciated in previous studies: the 
majority ($\sim$80\%) of randomly chosen disk galaxies do {\it not}\ have 
well-behaved velocity fields suitable for gas-dynamical modeling at \hst\ 
resolution.  Based on this experience, we have developed an efficient strategy 
to preselect galaxies that significantly increases the chance of success for 
future STIS programs.  This paper describes our technique, which is based on 
the use of the dust morphology to predict the velocity field.

\clearpage
\begin{figure*}[t]
\figurenum{1{\it a}}
\centerline{\psfig{file=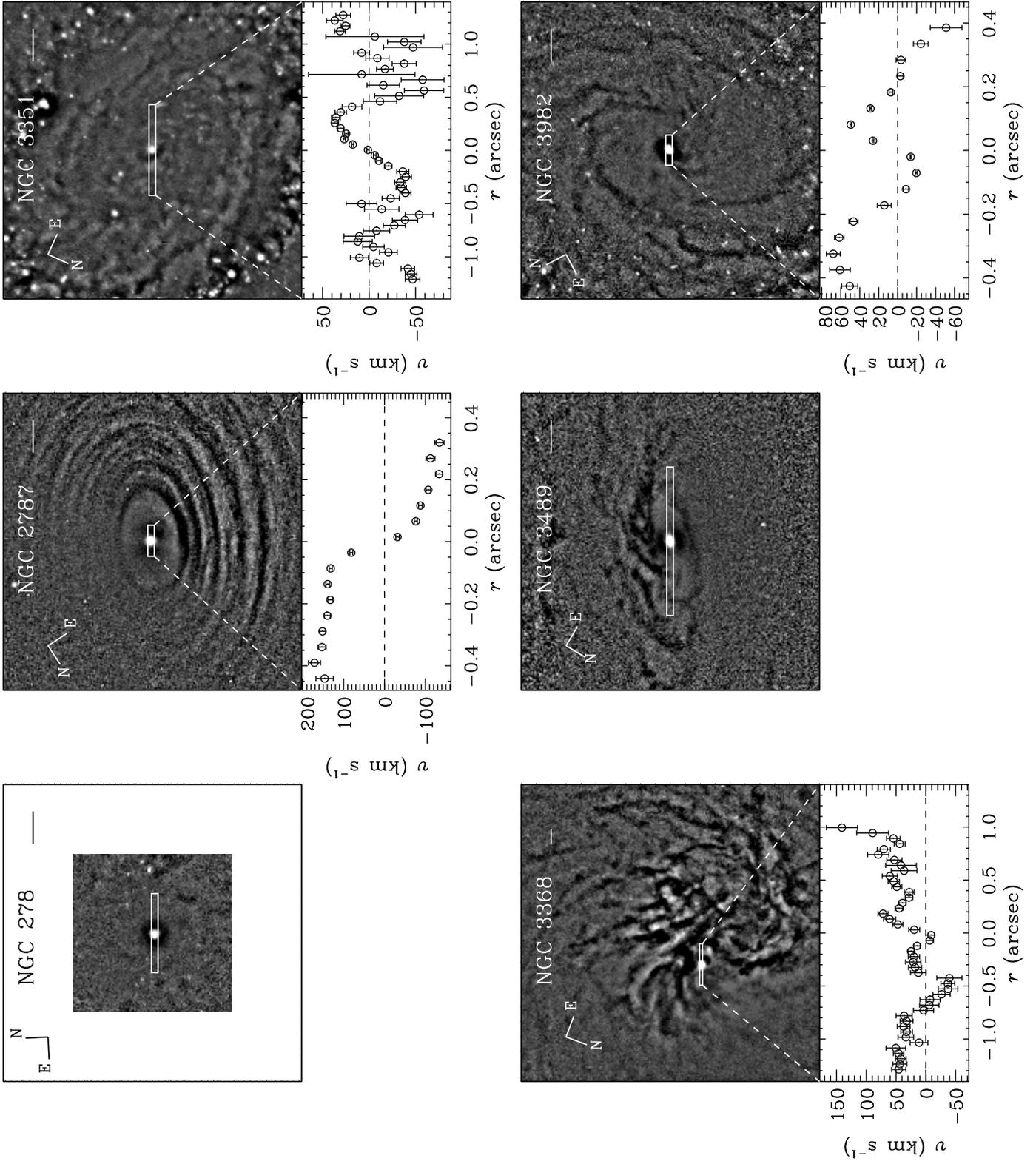,width=17.0cm,angle=180}}
\caption{
Dust morphology and gas kinematics for NGC 278, 2787, 3351, 3368, 3489, and
3982.  For each galaxy, the {\it top}\ panel shows the unsharp-masked, 
central 9\farcs1$\times$9\farcs1 region of either a $V$-band WFPC2 image or 
the F28X50LP STIS acquisition image.  (For NGC 3368, the image is a
20\asec$\times$20\asec\ section of the WF2 chip.) The horizontal bar on the
upper right corner gives the scale of 1\asec.  The rectangular region 
superposed on the nucleus outlines the position of the slit.  The 
{\it bottom}\ panel shows the H$\alpha$ velocity curve of the central 
$\sim$0\farcs5--1\farcs0, where the rest-frame velocity of the galaxy was 
determined by centering the rotation curve on the peak of the stellar 
continuum.  The velocity curve is omitted for galaxies lacking extended 
emission.
}
\end{figure*}
 
\clearpage
\begin{figure*}[t]
\figurenum{1{\it b}}
\centerline{\psfig{file=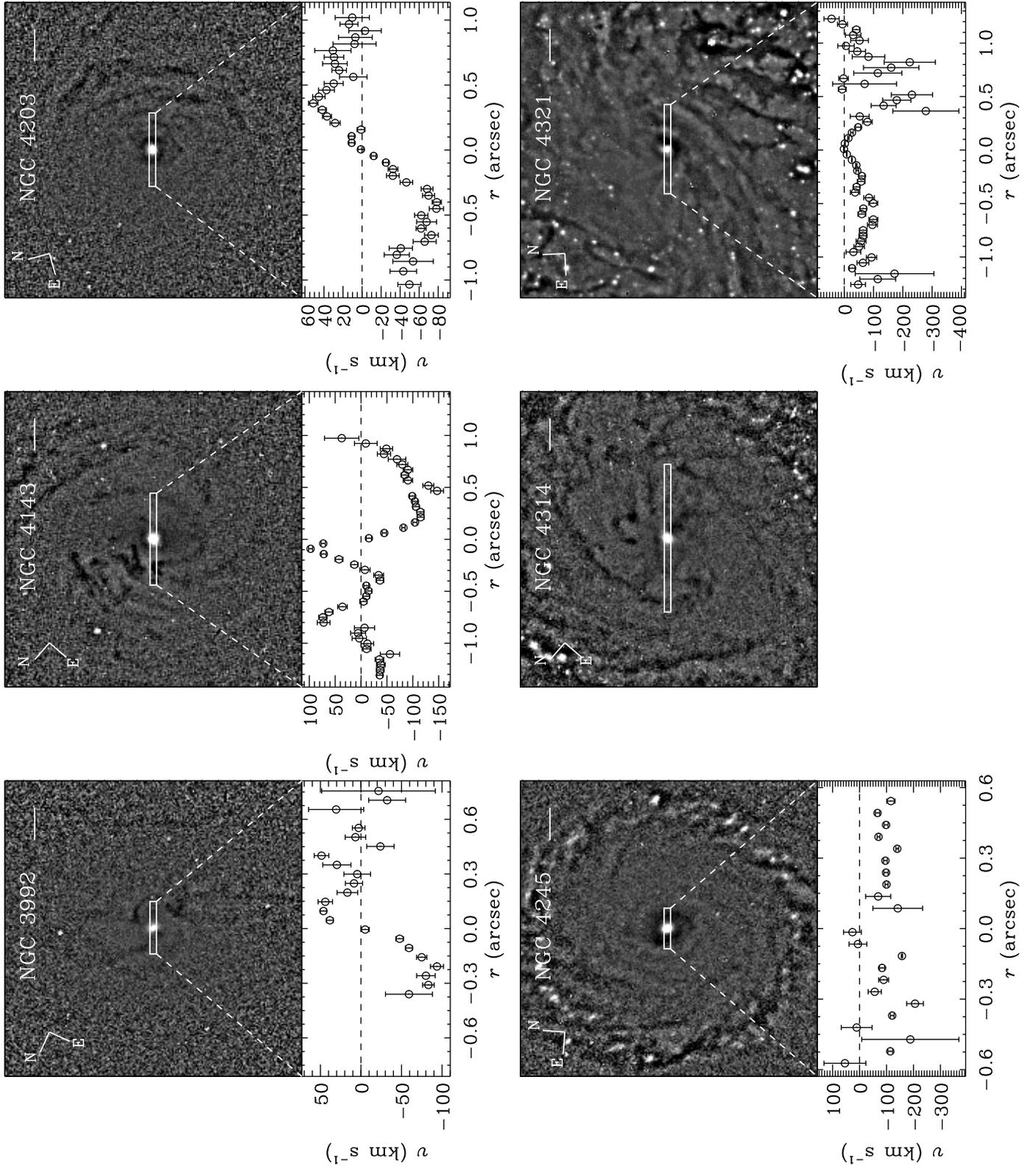,width=17.0cm,angle=180}}
\caption{
Dust morphology and gas kinematics for NGC 3992, 4143, 4203, 4245, 4314, and
4321.  As in Figure~1{\it a}.
}
\end{figure*}
 
\clearpage
\begin{figure*}[t]
\figurenum{1{\it c}}
\centerline{\psfig{file=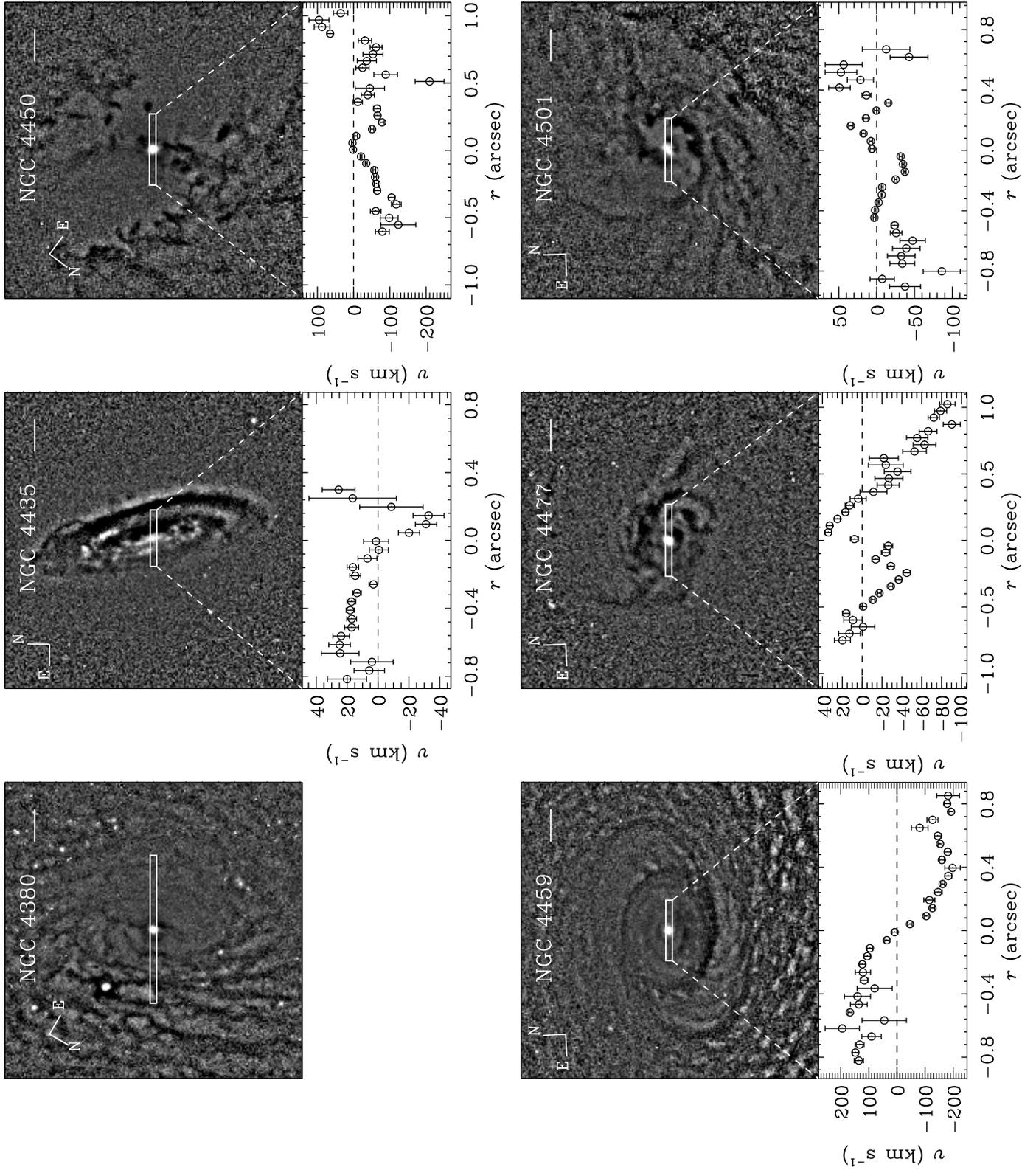,width=17.0cm,angle=180}}
\caption{                
Dust morphology and gas kinematics for NGC 4380, 4435, 4450, 4459, 4477, and
4501.  As in Figure~1{\it a}.
}
\end{figure*}
 
\clearpage
\begin{figure*}[t]
\figurenum{1{\it d}}
\centerline{\psfig{file=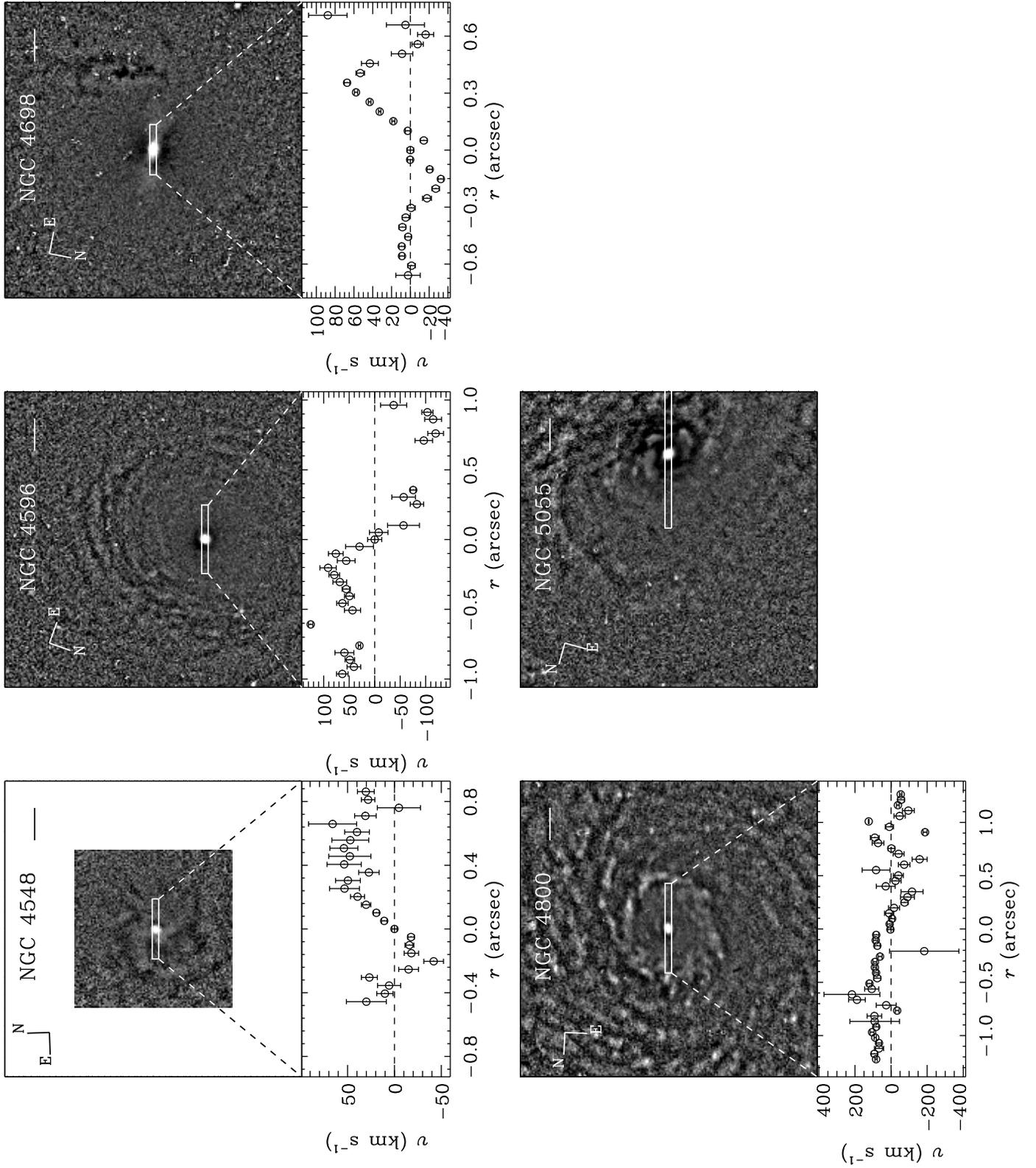,width=17.0cm,angle=180}}
\caption{
Dust morphology and gas kinematics for NGC 4548, 4596, 4698, 4800, and 5055.
As in Figure~1{\it a}.
}
\end{figure*}

\clearpage



\psfig{file=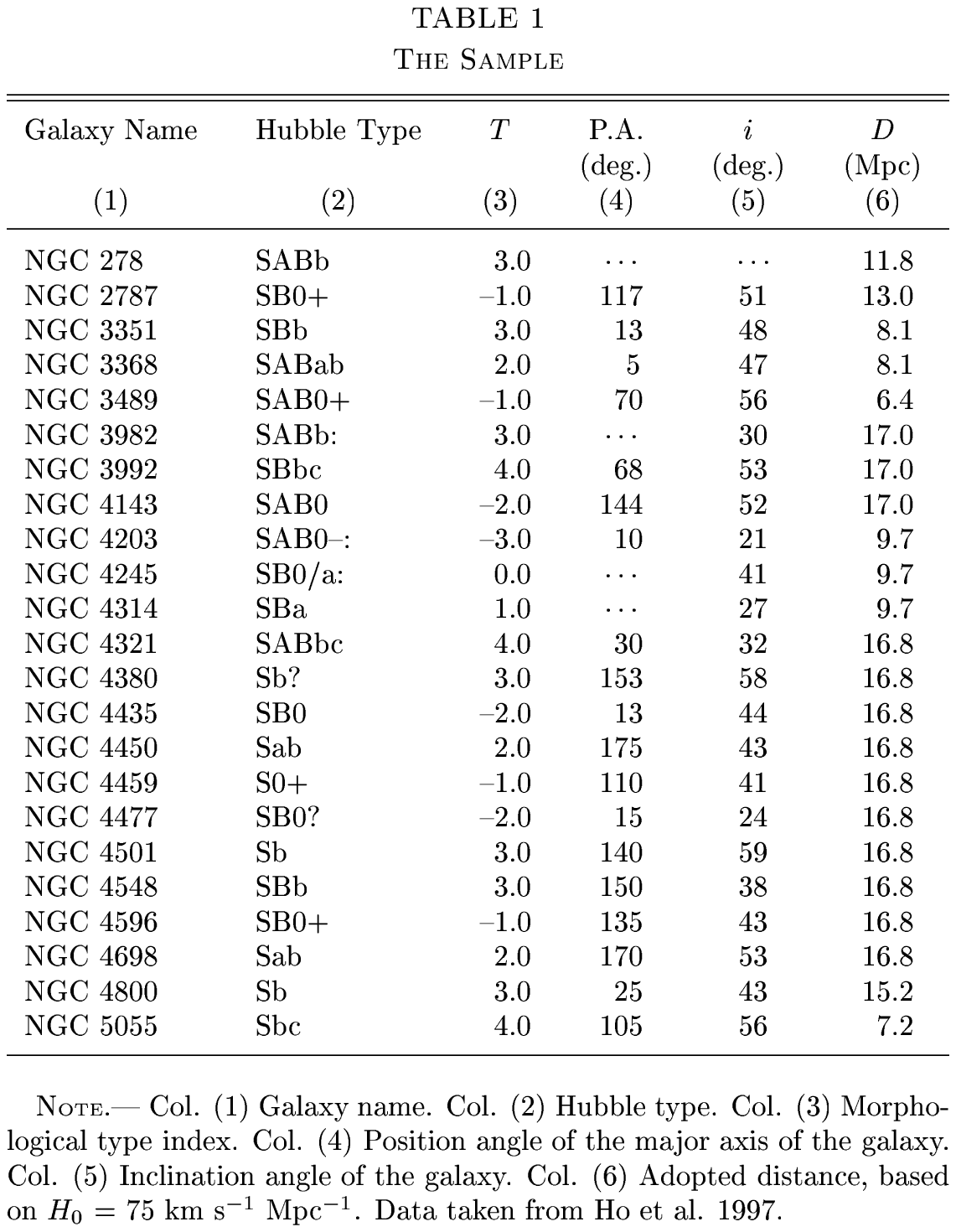,width=9.5cm,angle=0}


\section{The Data}

The data used in this paper derive from our Cycle~7 STIS program GO-7361.
Since a full description of the observations and data reductions will be given
elsewhere (Rix et al. 2002; see also Sarzi et al. 2001, 2002), here we only
mention a few key points pertinent to the present discussion.  Our sample was
designed to address a diverse set of scientific goals, among them measurement
of BH masses in bulges.  It comprises nearly all northern, bright S0--Sb
galaxies within 17~Mpc that are known to have nuclear emission in the
spectroscopic catalog of Ho, Filippenko, \& Sargent (1997).  We avoided
galaxies later than Sb because their bulges are expected to be faint or 
absent, and because their higher dust content may compromise our ability to 
center the narrow slit on the nucleus.  To further ensure that the nucleus 
will be easily accessible, we eliminated galaxies more inclined than 60\deg.  
The visibility of the nucleus was also confirmed by inspection of WFPC2 
images, which were available for most of the galaxies.   The final sample 
contained 24 galaxies, of which 23 (Table~1) were successfully observed; 
the acquisition of NGC 4138 failed because a nearby, bright star was 
mistaken for the nucleus.

Following a brief target-acquisition image using the F28X50LP long-pass 
filter, the primary kinematics data for each galaxy consist of a single-orbit 
integration with the G750M grating (\lamb\lamb 6300--6870 \AA; FWHM $\Delta 
\lambda \approx 2.2$ \AA) using the 0\farcs2$\times$52\asec\ slit.  To enable 
flexibility in scheduling, we did not choose specific position angles for 
the slit.  

\vspace*{0.5cm}

\section{Dust Morphology and Velocity Field}

We detected H\al\ and \nii\ \lamb\lamb 6548, 6583 emission in all the objects, 
but the emission was symmetrically extended in the central 0\farcs5--1\farcs0 
in only $\sim$75\% (17/23) of the sample.   For these 17 objects, ``rotation 
curves'' were derived from the radial velocities of the gas as a function of 
position along the slit using the H$\alpha$ and \nii\ emission lines.  The 
velocities of these features were derived using SPECFIT (Kriss 1994) within 
IRAF\footnote{IRAF is distributed by the National Optical Astronomy 
Observatories, which are operated by the Association of Universities for 
Research in Astronomy, Inc., under cooperative agreement with the National 
Science Foundation.}, assuming a common velocity, a Gaussian profile, and a 
fixed \nii\ doublet ratio of 3:1.   The results are shown in Figure~1.  
A striking result is the rarity of galaxies with well-behaved velocity 
fields. Among the 17 galaxies with spatially resolved velocity profiles, 
only four (NGC 2787, 4203, 4459, 4596), or 25\%, show ordered rotation curves 
amenable to dynamical analysis (Sarzi et al. 2001).  The rest of the velocity 
curves look disorganized and difficult to interpret (Fig.~1).  In terms of 
using the data for BH searches, therefore, the success rate is 
disappointingly low.

The greyscale panels in Figure~1 display the unsharp-masked optical continuum 
image of each galaxy.  Most of the data are archival broad-band (roughly $V$) 
images acquired with the PC detector on WFPC2; two are our own STIS F28X50LP 
target-acquisition images.  To accentuate high-spatial frequency features such 
as star clusters and dust lanes, we subtracted from each image a version of 
itself smoothed by a Gaussian with $\sigma$ = 4 pixels ($\sim$0\farcs2).  The 
nucleus, by design, is clearly visible in each image.  As was known from 
previous \hst\ imaging studies of disk galaxies (Phillips et al. 1996; Carollo 
et al. 1997), dust is ubiquitous.  Every galaxy in our sample shows dust lanes 
at some level within a central radius of $\sim$5\asec, or $\sim$400 pc for a 
typical distance of 17 Mpc.  However, unlike the case of early-type (E and S0) 
galaxies (van~Dokkum \& Franx 1995; Tomita et al. 2000; Tran et al. 2001), the 
dust distributions do not resemble simple disks with sharply delineated, outer 
edges.  Instead, the dust morphology can be best described as nuclear spirals 
(e.g., NGC 3351, NGC 3982), concentric rings (e.g., NGC 2787, NGC 4596), or 
simply chaotic (e.g., NGC 3368, NGC 4450).  The one object with a fairly 
well-defined nuclear disk, NGC 4435, turns out to be an S0 galaxy.

Closer examination of the data reveals that the degree of "orderliness" of the
gas kinematics evidently correlates with the morphology of the dust lanes. All 
four galaxies with well-behaved velocity fields (NGC 2787, 4203, 4459, 4596) 
have relatively smooth, circularly symmetric dust rings.  The converse seems 
to hold for most, but not all, objects.  For example, the dust patterns in 
NGC 4245 and NGC 4800 resemble those in the above-mentioned four objects, but 
neither has a regular, symmetric velocity field in ionized gas.  The case of 
NGC 4435 deserves special remark.  Although this galaxy possesses a clean 
nuclear disk, we were unable to obtain a usable symmetric rotation curve 
because the slit fell near the minor axis of its highly inclined disk 
(Fig.~1{\it c}).  The velocity curve of the far side of the disk indeed
looks regular, but that of the near side is obstructed by a prominent dust
lane.

Thus, in terms of planning for \hst\ BH searches using gas kinematics, it 
appears that {\it the morphology of the dust lanes can be used as an efficient 
and  relatively reliable predictor of the gas kinematics.}  Albeit somewhat 
subjective, and by no means perfectly foolproof, the strategy we advocate is 
nonetheless a highly effective tool for culling promising targets for STIS 
spectroscopy.  It is especially attractive in view of the extensive database 
of optical images of nearby galaxies available in the \hst\ archives and the 
straightforward manner in which unsharp masking can be performed on them.  
It would be fruitful to revisit these issues in the future when we have 
available images and spectra with resolutions higher than that currently 
possible with {\it HST}.

\section{Summary}

Searches for massive BHs in bulges of disk galaxies generally must rely on 
gas-dynamical methods because these objects are dusty. Our experience with 
STIS data indicates that the success rate will be low in randomly chosen 
samples of spiral and S0 galaxies because well-defined nuclear gas disks with 
regular, symmetric velocity fields are rare in these objects. We show that 
the degree of regularity in the velocity field correlates with the morphology 
of the dust lanes, which can be efficiently ascertained through unsharp 
masking of optical continuum \hst\ images.  Objects with smooth, circularly 
symmetric dust lanes tend to have velocity fields that are amenable to 
dynamical analysis.  By contrast, galaxies with obviously disorganized dust 
morphology will almost never yield easily interpretable velocity fields.  We 
draw attention to this empirical correlation as a powerful tool to select 
targets for future \hst\ programs to search for massive BHs using gas 
kinematics.

\acknowledgments
This work was supported financially through NASA grant NAG 5-3556, and by 
grants in support of \hst\ programs GO-7361 and GO-7403, awarded by STScI, 
which is operated by AURA, Inc., for NASA under contract NAS5-26555. Research 
by A.~J.~B. is supported by a postdoctoral fellowship from the 
Harvard-Smithsonian Center for Astrophysics. A.~V.~F. is grateful for a 
Guggenheim Foundation Fellowship."


%
%
%
%

\end{document}